\documentclass[12pt,preprint]{aastex}

\shorttitle{Curved HH Jets}
\shortauthors{Ciardi et al.}

\begin{document}


\title{Curved Herbig-Haro Jets: Simulations and Experiments}

\author{A. Ciardi}
\affil{Observatoire de Paris, LERMA, 92195, Meudon, France.}
\email{andrea.ciardi@obspm.fr}

\author{D.J. Ampleford}
\affil{Sandia National Laboratories, Albuquerque, NM 87123-1106, USA.}
\email{damplef@sandia.gov}

\author{S.V. Lebedev}
\affil{Blackett Laboratory, Imperial College, London SW7 2BW, United Kingdom.}
\email{s.lebedev@imperial.ac.uk }

\and

\author{C.Stehle}
\affil{Observatoire de Paris, LERMA, 92195, Meudon, France.}
\email{chantal.stehle@obspm.fr}

\begin{abstract}
Herbig-Haro jets often show some degree of curvature along their path, in many cases produced by the ram pressure of a side-wind. We present simulations of both laboratory and astrophysical curved jets and experimental results from laboratory experiments. We discuss the properties and similarities of the laboratory and astrophysical flow, which show the formation of internal shocks and working surfaces. In particular the results illustrate how the break-up of the bow-shock and clumps in the flow are produced without invoking jet variability; we also discuss how jet rotation reduces the growth of the Rayleigh-Taylor instability in curved jets.
\end{abstract}


\section{Introduction}
Parsec-scale Herbig-Haro (HH) jets are one of the most spectacular phenomena connected to star formation. Observations of line emission features reveal an intricate succession of bow shocks and knots along the jet, and a complex and rich dynamics arising from the interaction with the circumstellar environment (see \citet{Reipurth01} for a review). In the reflection nebula NGC 1333, in the Perseus molecular cloud, a number of bipolar HH jets exhibit a distinguishing C-shape morphology indicative of a steady bending \citep{Bally01}. Less regular curvature is also observed in a number of other HH jets; for example in HH 30 a small side drift close to the jet source is followed further away by a sudden bending \citep{Anglada07}. In general, the curvature in jets has been linked either with the motion of the jet sources relative to the ambient medium or with the presence of a widespread outflow; both cases giving rise to an effective transverse ram pressure (cross-wind) which curves the jet. In fact, numerical simulations of jet bending produced by magnetic fields seem to require field strength that are unrealistically high \citep{Hurka99}. For the jets in the Orion Nebula, it was also suggested that the bending may be caused by the 'rocket' effect induced by the photo-ionizing UV radiation from nearby high-mass stars \citep{Bally01}; recent work by \citet{Kajdic07} has investigated the influence of an ionizing photon flux on the jet-wind collision in the HH555 jet. The presence of curved jets is also common to other environments, such as in active galactic nuclei where they are observed as narrow-angle-tail radio sources (O'Dea and Owen 1986), or the pulsar jets in the Vela nebula, which are curved under the combined action of the wind within the supernovae remnant and the proper motion of the pulsar (Pavlov, Teter et al. 2003).
Under simplifying assumption the curvature of the jets can be described analytically \citep{Begelman79, Canto95}, however these models cannot detail the shock structures and the intricate dynamics developing in the interaction, and that were observed in  a number of numerical studies. For the narrow-angle-tail sources \citep{Balsara92} internal shocks were shown to be responsible for the formation of density enhancements (knots) inside the jet beam which was then disrupted through the combined effect of Kelvin-Helmholtz and Rayleigh-Taylor instabilities. HH jets were first investigated numerically by \cite{Lim98} for different incident angles of the wind and by \citet{Masciadri01} for the case of irradiated jets. More recently the jet-wind collisions have also been studied in the laboratory \citep{Lebedev04, Ampleford07} with supersonic (Mach number $\sim20-40$) jets interacting with a fast, radiatively ablated plasma that acted as a cross-wind. The laboratory results clearly demonstrated a steady curvature in the jets under the impinging cross-wind and the appearance of internal shocks.

	In the present work we combine consistently the laboratory modelling (experiments and simulations) with simulations of curved HH jets, drawing a parallel between the observed dynamics in the two systems and, with in mind future experiments, we also discuss the effects of rotation on curved jets.

\section{Laboratory Jets}
Laboratory jets are produced on the MAGPIE pulsed-power generator \citep{Mitchell96} using a conical cage of micron-sized metallic wires (Al, Fe, W, etc) which is subject to a 1 Mega Ampere current, with a rise-time of 250 ns (see \citet{Lebedev05} for a review). Figure 1 shows a schematic of the experimental set-up; the ohmically heated wires rapidly vaporize and turn into streams of plasma accelerated by the \textbf{j}$\times$\textbf{B}$_G$ force towards the axis of the array. The collision of the streams forms a standing conical shock, which redirects the plasma axially into a jet. Radiative cooling rapidly reduces the temperature in the jet which attains high Mach numbers (M$\sim20-40$) and small opening angles. The interaction of the jet with a side-wind is replicated in the laboratory by placing a plastic foil in the jet propagation region. XUV radiation from the standing shock radiatively ablates the foil and generates a fast moving plasma ($\sim30-50$ km s$^{-1}$), providing the ram pressure to bend the jet. By changing the position, angle and length of the foil the characteristic parameters of the impinging wind can be altered, and different regimes of the interaction studied \citep{Lebedev04, Ampleford07}. Similar to the HH jets, the characteristic cooling time in the laboratory jets is much smaller than the hydrodynamic time, this ratio is $\geq0.1$, and radiation losses play a significant role in the energy balance of the jets. The numerical simulations are performed with the three-dimensional resistive MHD code GORGON \citep{Chittenden04, Ciardi07}, the ion and electron energies are solved separately and include electron and ion thermal conductions. The coupling of the two energy equations due to collisional heat exchange is included as a sink/source term. The plasma is assumed to be in local thermodynamic equilibrium (LTE) and the average ionization is calculated by a Thomas-Fermi model. The plasma is optically thin and a radiation losses term is included in the electron energy equation.

	Two-dimensional axis-symmetric MHD simulations of the whole conical wire array were performed to obtain the mass and momentum fluxes of the jet. The calculated time and spatially varying profiles are then used as boundary conditions to inject the jet into the three-dimensional (3D) computational domain. Typical axial jet velocities and ion densities vary in the range $v_j\sim100-200$ km s$^{-1}$ and $n_j\sim10^{18}-10^{20}$ cm$^{-3}$ respectively  \citep{Ciardi02}; the jet is injected with an initial temperature $T_j=15$ eV $\sim175\times10^{3}$ K. The material of the wires, and thus the jet, is tungsten. In the jet and interaction region the magnetic field is negligible and in the 3D simulations presented here only the gasdynamic equations are integrated in time. The physical conditions of the cross-wind, produced by photo-ablation of the foil, depend on the incident radiation flux coming from the conical shock and are also time-dependent. However the simulations show that within the range of the expected and measured wind parameters \citep{Ampleford07}, the dynamics of the interaction remains substantially the same and a detailed treatment of the time-variation of the wind is not essential. For the present simulations the injected wind has a constant velocity ($v_w=60$ km s$^{-1}$) and temperature ($T_j=30$ eV), the characteristic ratio of the jet velocity (which we remind is time-dependent) to the wind velocity $v_j/v_w\sim2-4$. Two wind density profiles are discussed, one is constant $n_w=1.2\times10^{19}$ cm$^{-3}$  (case A) and the other is exponentially increasing (case B). However for both cases the density contrast ratio is in the range $n_j/n_w\sim0.1-10$; the wind is supersonic and pressure gradients play no significant role in bending the jet.
	
	Simulation results for the laboratory jet-wind interaction are shown in Figure 2 as XUV emission maps (photon energies $h\nu>30$ eV) integrated along the line of sight. The emitted radiation includes bound-bound, free-bound and free-free transition and it is calculated assuming LTE within a screened hydrogenic model \citep{Stehle02, Michaut04}. Self-absorption of radiation and scattering are not taken into consideration. In the images the jet is injected through the lower boundary and the wind is injected on the left boundary. The collision of the wind with the jet generates a bow shock which envelopes the whole jet. As the bow shock is advected downstream with the wind, it develops a highly asymmetric shape, with the upwind side of the jet cocoon effectively disappearing. An oblique shock forms in the jet body which begins to bend and redirect the jet flow sideways. It is clear that the jet bends considerably more in case A where the time-averaged ram pressure of the wind is about an order magnitude higher than in case B. As the jet curves, momentum transferred from the jet to the working surface rapidly decreases, until the working surface effectively detaches from the jet; a result described by \citep{Masciadri01} in the context of simulations of the jet HH 505. The subsequent propagation of the working surface is ballistic and as it entrains more ambient mass it slows down and its emission rapidly decays. The presence of internal shocks in the jet beam is visible in both the simulations and experiments. The experimental XUV self-emission image (Figure 3) shows the formation of a new working surface half-way up the image and enhanced emission inside the jet beam. These shocks are produced by perturbations developing along the jet-wind interface and their effect is to slow down and redirect the jet sideway, allowing the un-shocked flow ahead to detach. The effect is particularly pronounced for the case, considered here, where the jet momentum increases in time. This produces a narrowing of the angle of the main oblique shock (with respect to the z-axis) and causes the jet to bore a new, straighter channel in the ambient medium. Also a time-varying wind would have the same effect. Nevertheless the formation of new working surfaces was seen in simulations of uniform laboratory jets and wind, and in the simulations of astrophysical jets presented next.

\section{Jets from young stars}
The relevance of laboratory experiments to astrophysical jets, and for the case discussed here the jet-wind interaction, rests on the ability of producing both an adequate representation of the dynamics of the astrophysical system and, more importantly, obtaining dimensionless parameters in the appropriate astrophysical range \citep{Ryutov99}. For the present experiments such conditions are well met \citep{Lebedev05} and we expect much of the dynamics observed in the laboratory to be valid for astrophysical jets.

Expected wind velocities vary from a few km s$^{-1}$ for the jet-wind interaction associated with relative motions of TTauri stars with respect to the surrounding environment, (see for example \citet{Jones79}) to typically higher velocities for irradiated jets: best fits to HH505 H$\alpha$ emission maps were obtained by \citet{Masciadri01} for a wind velocity of 15 km s$^{-1}$ and estimates by \citet{Bally01} give wind velocities in the Orion nebula and in NGC1333 of $\sim10-20$ km s$^{-1}$. Here we present simulations of HH jets in a cross-wind but do not model a specific jet. Instead we maintain the ratios of jet and wind velocities, and densities in a similar range as those obtained in the laboratory system. Specifically we take for the jet $v_j=100$ km s$^{-1}$, $n_j=1000$ cm$^{-3}$ and for the wind $v_w=25$ km s$^{-1}$, $n_w=100$ cm$^{-3}$. Although these values are consistent with observations, the somewhat larger wind velocity was chosen to produce enough bending in the jet while at the same time maintaining as much as possible of the interesting dynamics within the computational domain. Lower $v_w$ produce qualitatively similar results but because of the smaller curvature radius require considerably more computational resources. The initial jet radius $r_j=50$ AU and the resolution is 4 AU over the entire Cartesian grid. The initial temperature in both jet and wind is 5000 K. The HH jet simulations are performed with an appropriately modified version of our laboratory code, where we now follow the time-dependent ionization of hydrogen. We take into account the recombination and collisional ionization and use the rate coefficients as tabulated in \citet{Raga07} (and references therein). For temperatures above 15000 K cooling is implemented by a function appropriate for interstellar gas composition \citep{Dalgarno72}. However for temperatures below 15000 K cooling is calculated by including collisional excitation, collisional ionization and radiative recombination of hydrogen, and the collisional excitation of O I and O II. Because of charge exchange the neutral and singly ionized populations of oxygen are assumed to follow closely those of hydrogen \citep{Hartigan93}; the atomic abundance of oxygen is taken to be $5\times10^{-4}$ that of hydrogen.
	
	The HH jet evolution is shown in Figure 4 in a time series of column density maps spanning 500 years of evolution. In general the overall jet dynamics is similar to that seen in laboratory jets. The presence of internal oblique shocks in the jet is apparent both in Figure 4 and also in Figure 5c which plots the H$\alpha$ emission. The simulations illustrate the formation of a number of bow-shaped shocks and knots in the resulting flow. It is worth noting that the clumpiness seen in the jet and the break-up of the bow shock arise entirely from the interaction with the cross-wind and are not imposed by means of a variable jet injection velocity. The latter is often used in numerical modeling to reproduce the knotty emissions features observed in the jets and which are thought to be caused by small velocity variations in the flow. The present simulations and more importantly the experiments, suggest that these internal shocks can also be produced by a wind impacting on a jet and are a dynamical observable feature dependent on the environment and not on the jet velocity profile. In addition to a steady cross-wind, the relative motion of the jet with respect to the interstellar medium, such as due to jet precession or the localized presence of transverse winds, may be also responsible for the formation of some of the knots observed. The detachment of the main working surface, as the jet body is increasingly curved under the action of the wind, occurs during the first 200 years and it is particularly evident in Figure 5b, which shows a passive tracer of the jet material at 370 years. The contact discontinuity between the jet and wind plasmas clearly shows perturbations along its surface which are ultimately responsible for producing the internal shocks in the jet. It is also evident that the jet is 'naked': the upstream side of the jet is essentially in contact with the wind and the cocoon is absent. As pointed out in the work of \citet{Balsara92} the jet is liable to the growth of the Rayleigh-Taylor instability (RTI). We find that at the interface between the jet and the shocked wind, density and pressure gradients are in opposite directions, also the effective acceleration experienced by the jet material is in the direction of the centrifugal force and it is pointing into the density gradient, making the interface unstable. The development of the RTI in the jet beam is shown in Figures 5e and 5d which plot slices across the computational box (\textit{x-y} plane) of the atomic number density. The wind is coming from the left and the jet is traveling out of the page. The arrows on the left-most panel indicate the height where the slices were taken.  A well defined bow shock around the jet is clearly visible and so is the shock internal to the jet on the upwind side (Figure 5e). This is where the RTI can be seen to grow. The initial perturbation develops the characteristic bubble and spike structure (Figure 5d), with the tip of spikes rolling-up on the sides due to Kelvin-Helmholtz instability (KHI). The RTI tends to split the jet into well defined filaments; in addition as it penetrates deep into the jet the peak emission inside the jet shifts to the down-stream side, resulting into a broadening of the emission across the whole diameter. However we might expect that any departures from uniform jet, wind and symmetry would modify the dynamics of the RTI.
	
	An example of particular interest is given by the simulation of a rotating jet shown in Figures 5f-j. The jet is taken to rotate as a solid body and we set the azimuthal velocity at its boundary $v_\phi=0.5c_s$, where $c_s$ is the initial sound speed in the jet. The other parameters of the wind and jet are as those of the non-rotating case. The evolution is indeed similar, however the cross section of the rotating jet is somewhat larger due to the increased radial expansion and the density is correspondingly lower. The result is that the rotating jet tends to have a higher curvature (see for example Figures 5b and 5g) and a shorter propagation length. The most interesting difference though, can be seen in the slices across the computational domain (Figure 5i and 5j). Because of rotation (anti-clockwise in the images) perturbations driven by the RTI are advected around the surface of the jet and are rapidly sheared seeding the KHI; however for the instabilities are confined to a narrower region of the jet. The H$\alpha$ emission maps in Figure 5c and 5h show that the internal shock for the rotating jet case remains narrower and the flow laminar for much longer. Thus rotation, combined with the action of the cross wind, reduces the growth of the RTI and the disruption of the jet. For the simulations presented here, the rotating jet survives up to $z\sim2000$ AU as opposed to $z\sim1400$ AU for the non-rotating case, above those heights the inabilities have grown to cover the whole cross sectional area of the jet.

\section{Discussion}
This work illustrates how combining laboratory experiments, numerical simulations of these experiments and the simulations of astrophysical systems can offer significant insights into the phenomena studied. In the case of the interaction of HH jets with a cross-wind, the results show the presence of oblique shocks and knots in curved jets, the formation of working surfaces, the detachment and fragmentation of the bow shock. In particular some of the observed features in HH jets may arise as a consequence of the curvature and the subsequent instabilities developing in the jet. However, it remains an open question what would be the minimum curvature required to produce them, especially for large values of the ratio $\rho_jv_j/\rho_wv_w$, where computations become very demanding. The simulations show the development of the RTI in curved jets and how it may be partially quenched by rotation, which rapidly shears the RT modes. To determine the observational signatures of such effect and for direct comparison to observation, more realistic angular velocity profiles need to be used (see for example \citet{Lery00}).
	
	A simple estimate of the RTI growth rate is obtained by assuming the plasma to be incompressible, its classical value is then given by $\omega=\sqrt{Akg}$ where $k$ is the wave number of the perturbation and $A=\left(\tilde{\rho_j}-\tilde{\rho_w}\right)/\left(\tilde{\rho_j}+\tilde{\rho_w}\right)$ is the Atwood number \citep{Chandrasekhar61} evaluated with the local densities across the interface. For the jets we take the effective acceleration $g$ on the jet fluid to be equal to the centrifugal acceleration:  $g=\kappa v^{2}_{j}$, where $\kappa=1/R$ and $R$ is the radius of curvature. Thus for the simulated HH jets we have: $v_j\sim10^5$ m s$^{-1}$, $R\sim3000$ AU, $k=2\pi/r_j\sim8\times10^{-13}$ m$^{-1}$, $A\sim0.3$; where we have taken the wavelength of the perturbation to be the jet radius. The characteristic growth time is then $\tau_{RTI}=\frac{2\pi}{\omega}\sim70$ years which compares very well with the growth time of the instability seen in the simulations. Laboratory jets also are susceptible to the RTI. However while it was seen in simulations, it is not clear if the filamentary structures observed in the experiments (cf. Figure 3) are caused by the growth of the RTI into the jet body. In general for the experimental jets $v_j\sim1.5\times10^5$ m s$^{-1}$, $R\sim0.05$ m, $k=6300$ m$^{-1}$, $A\sim1$; giving a characteristic growth time, $\tau_{RTI}\sim150$ ns, which is comparable to the propagation time of the jet in the cross-wind.
		
	A useful relation to estimate the growth rate of the RTI can be simply derived by using the analytical expression of the curvature radius at the stagnation point for an isothermal jet given in \citet{Canto95},  $R=r_jv_j^{2}v_w^{-1}c^{-1}_s(\rho_j/\rho_w)^{1/2}$. The growth rate can then be written as 
	\[\omega=(2\pi)^{1/2} A^{1/2}(\rho_w/\rho_j)^{1/4}(c_s v_w)^{1/2}r_j^{-1}
\]
which also shows that within these simplifying approximations $\omega$ does not depend on the jet velocity. The sound speed $c_s$ should be evaluated in the region of the jet body in pressure equilibrium with the shocked wind. Thus using reasonable values for the perturbed sound speed, $c_s\sim5-15$ km s$^{-1}$, we find for the simulated laboratory and astrophysical jets $\tau_{RTI}\sim150-250$ ns and $\tau_{RTI}\sim100-200$ years respectively; in good agreement with the previously estimated characteristic growth times. To observe the full development of the RTI in the laboratory jets we expect that experiments with longer interaction times ($>\tau_{RTI}$) will be required. Such work is planned for the future and it will be coupled with the study of the effects of rotation on the jet. In fact, by using a modification of the conical array described in the present work we have recently demonstrated \citep{Ampleford07b} the possibility of producing rotating jet in the laboratory with ratios of the axial to azimuthal velocity, $v_j/v_\phi\sim5-10$, of the same order as those reported in recent observation of HH jets \citep{Coffey04}.

\acknowledgments
The authors wish to thank Fabio de Colle and Sylvie Cabrit for useful discussions. The present work was supported in part by the European Community's Marie Curie Actions - Human Resource and Mobility within the JETSET (Jet Simulations Experiments and Theory) network under contract MRTN-CT-2004 005592. The authors also wish to acknowledge the London e-Science Centre (LESC) for the provision of computational facilities and support. 



\begin{figure}
\epsscale{.50}
\plotone{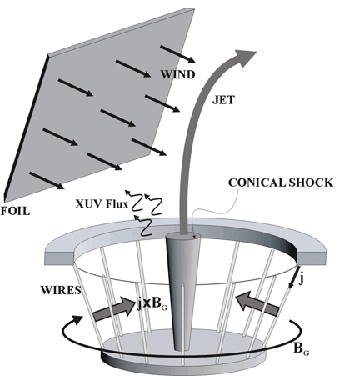}
\caption{Schematic of the experimental set-up.}
\label{fig1}
\end{figure}


\begin{figure}
\epsscale{0.5}
\plotone{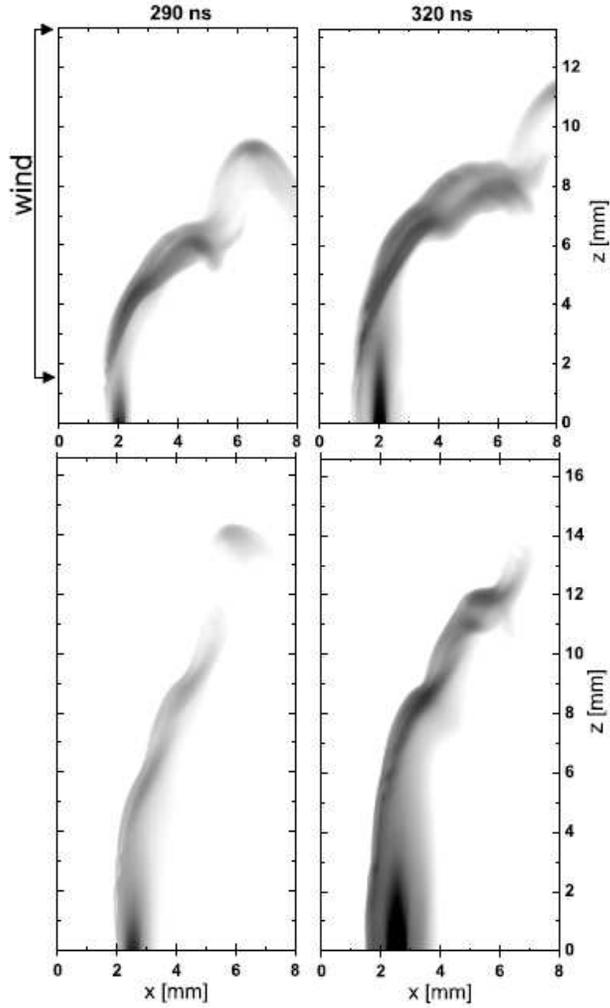}
\caption{Synthetic XUV emission images of the laboratory jets shown at two times 290 ns and 320 ns from the start of the current pulse. Top two panels for case A and bottom two panels for case B (see text). The wind is injected through the left boundary. For case A the wind injection length is indicated; for case B the wind injection is over the entire left boundary.}
\label{fig2}
\end{figure}

\begin{figure}
\epsscale{0.5}
\plotone{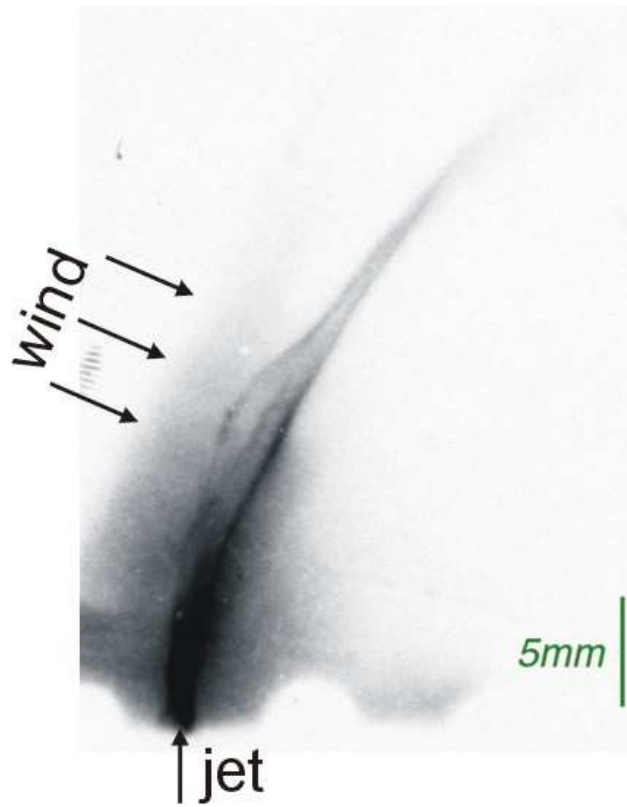}
\caption{Experimental time-resolved XUV image at 380 ns after the start of current. The foil in the experiments has an inclination of $30^o$ with respect to the z-axis.}
\label{fig3}
\end{figure}


\begin{figure}
\epsscale{.70}
\plotone{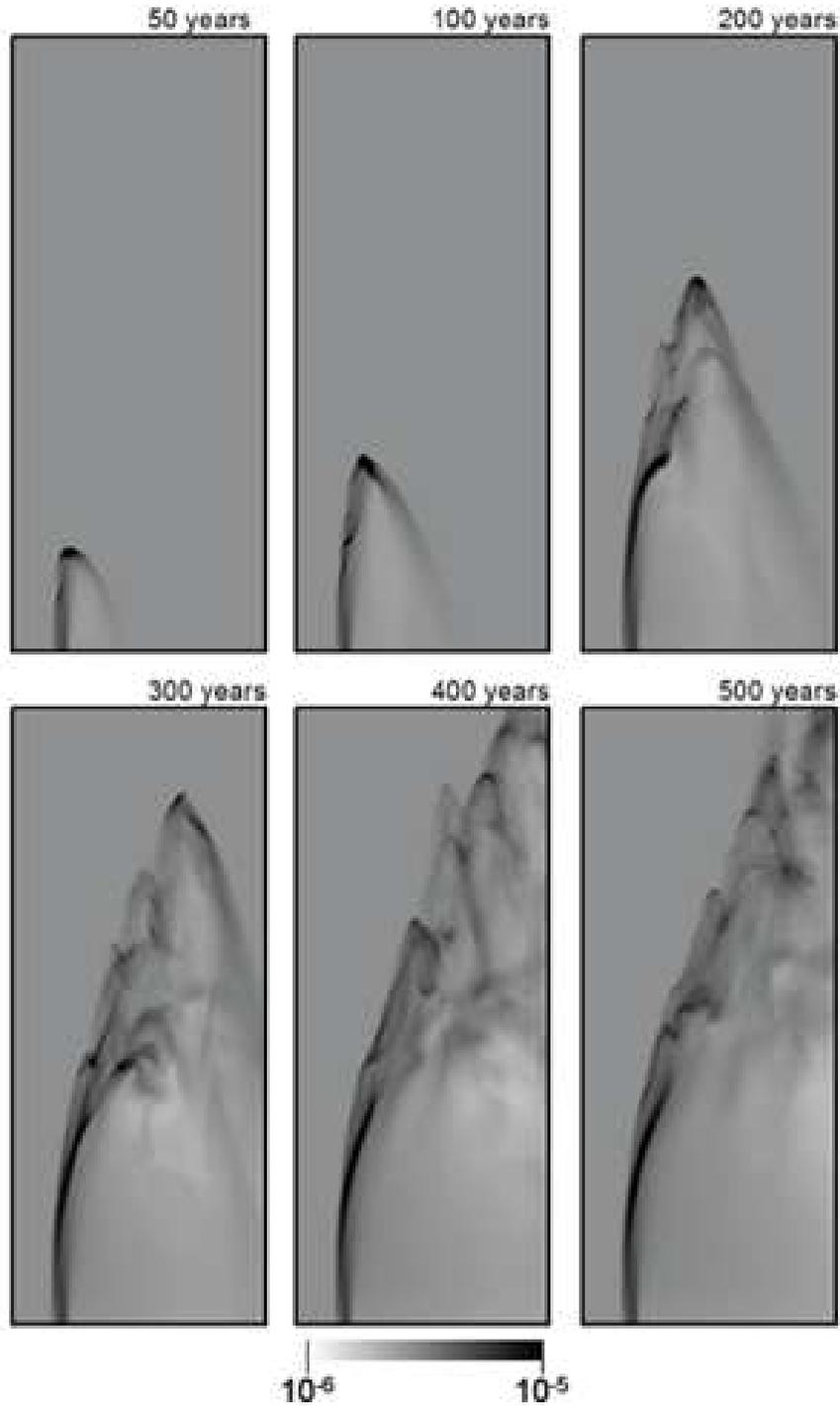}
\caption{Time evolution of jet-wind interaction for astrophysical condition relevant to YSOs jets. The plots show on a linear scale the column density (g cm$^{-2}$). The computational domain has a size $2004\times4864$ AU.}
\label{fig4}
\end{figure}


\begin{figure}
\epsscale{.60}
\plotone{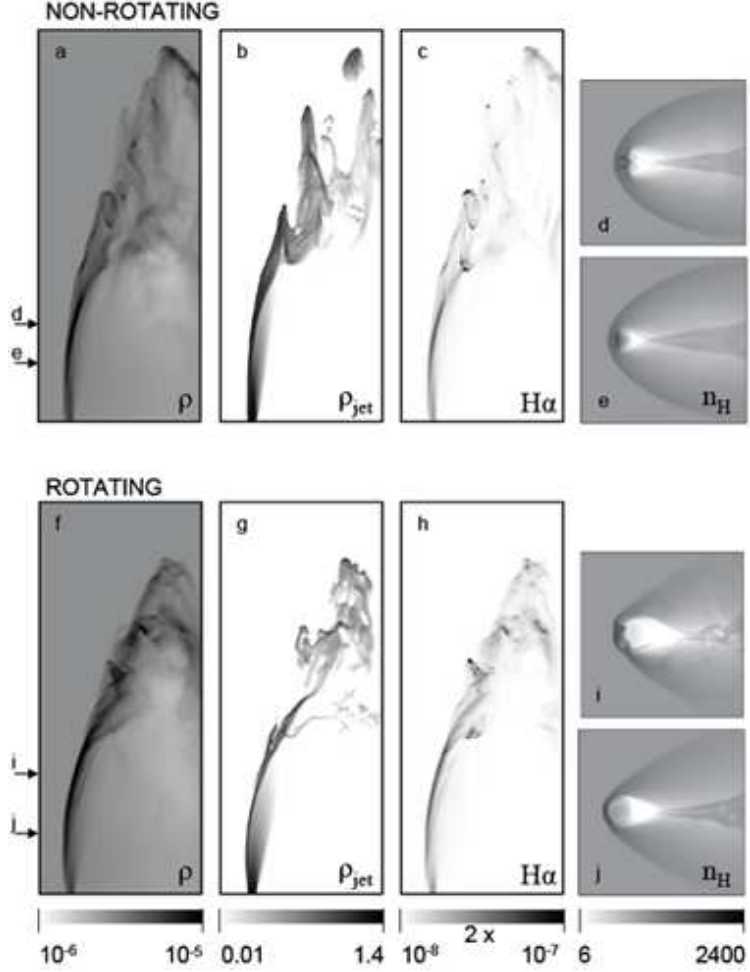}
\caption{Comparison of the jet-wind collision for the non-rotating jet (panels a-e) and the rotating jet (panels f-j) cases at 370 years. Panels a-c and f-h show slices in the \textit{x-z} plane of various quantities integrated along the y-direction; the size of the domain shown is $2004\times4864$ AU. The combined jet-wind column density (g cm$^{-2}$) is shown in (a) and (f). Panels (b) and (g) show a tracer of the jet plasma on a logarithmic scale in arbitrary units. Panels (c) and (h) plot the H$\alpha$ emission (erg cm$^{-2}$ s$^{-1}$ sr$^{-1}$) on a linear scale. Panels d, e, i and j are slices in the \textit{x-y} plane across the computational domain and show the hydrogen number density (cm$^{-3}$) on a logarithmic scale. The size of the domain shown is $2004\times2004$ AU. The heights where the slices were taken is indicated by the arrows on the left-most panels; panels e and j correspond to $z=760$ AU, panel d to $z=1261$ AU and panel i to $z=1520$ AU.}
\label{fig5}
\end{figure}


\begin{thebibliography}{}
\bibitem[Anglada et al. (2007)]{Anglada07} Anglada, G., L\'opez, R., Estalella, R., Masegosa, J., Riera, A. \& Raga, A. C. 2007, \aj , 133, 2799-2814
\bibitem[Ampleford et al. (2007)]{Ampleford07} Ampleford, D. J., Ciardi, A., Lebedev, S. V., Bland, S. N., Bott, S. C., Chittenden, J. P., Hall, G. N., Frank, A. et al. 2007, \apss , 307, 29-34
\bibitem[Ampleford et al. (2007b)]{Ampleford07b} Ampleford, D. J., Lebedev, S. V., Ciardi, A., Bland, S. N., Bott, S. C., Hall, G. N., Naz, N., Jennings, C. A. et al. 2007, \prl , submitted
\bibitem[Bally \& Reipurth (2001)]{Bally01} Bally, J. \& Reipurth, B. 2001, \apj , 546, 299-323
\bibitem[Balsara \& Norman (1992)]{Balsara92} Balsara, D. S. \& Norman, M. L. 1992, \apj , 393, 631-647
\bibitem[Begelman et al. (1979)]{Begelman79} Begelman, M. C., Rees, M. J. \& Blandford, R. D. 1979, \nat , 279, 770-773
\bibitem[Canto \& Raga (1995)]{Canto95} Canto, J. \& Raga, A. C. 1995, \mnras , 277 1120-1124
\bibitem[Chandrasekhar (1961)]{Chandrasekhar61} Chandrasekhar, S. 1961 International Series of Monographs on Physics, Oxford: Clarendon, 1961 
\bibitem[Chittenden et al. (2004)]{Chittenden04} Chittenden, J. P., Lebedev, S. V., Jennings, C. A., Bland, S. N. \& Ciardi, A. 2004 Plasma Physics and Controlled Fusion 46 B457-476
\bibitem[Ciardi et al. (2002)]{Ciardi02} Ciardi, A., Lebedev, S. V., Chittenden, J. P. \& Bland, S. N. 2002, Laser and Particle Beams, 20, 2, 255-262C
\bibitem[Ciardi et al. (2007)]{Ciardi07} Ciardi, A., Lebedev, S. V., Frank, A., Blackman, E. G., Chittenden, J. P., Jennings, C. J., Ampleford, D. J., Bland, S. N. et al. 2007, Physics of Plasmas, 14, 056501
\bibitem[Coffey et al. (2004)]{Coffey04} Coffey, D., Bacciotti, F., Woitas, J., Ray, T. P. \& Eislöffel, J. 2004, \apj , 604, 758-765
\bibitem[Dalgarno \& McCray (1972)]{Dalgarno72} Dalgarno, A. \& McCray, R. A. 1972, \araa , 10, 375
\bibitem[Esquivel et al. (2007)]{Esquivel07} Esquivel, A., Raga, A. C. \& de Colle, F. 2007, \aap , 468, 613-616
\bibitem[Hartigan \& Raymond (1993)]{Hartigan93} Hartigan, P. \& Raymond, J. 1993, \apj ,409, 705-719
\bibitem[Hurka et al. (1999)]{Hurka99} Hurka, J. D., Schmid-Burgk, J. \& Hardee, P. E. 1999, \aap , 343, 558-570
\bibitem[Jones \& Herbig (1979)]{Jones79} Jones, B. F. \& Herbig, G. H. 1979, Astronomical Journal, 84, 1872-1889
\bibitem[Kajdi$\breve{c}$ \& Raga (2007)]{Kajdic07} Kajdi$\breve{c}$, P. \& Raga, A. C. 2007, \apj, 670, 2, 1173-1177
\bibitem[Lebedev et al. (2004)]{Lebedev04} Lebedev, S. V., Ampleford, D., Ciardi, A., Bland, S. N., Chittenden, J. P., Haines, M. G., Frank, A., Blackman, E. G. et al. 2004, \apj , 616, 988-997
\bibitem[Lebedev et al. (2005)]{Lebedev05} Lebedev, S. V., Ciardi, A., Ampleford, D. J., Bland, S. N., Bott, S. C., Chittenden, J. P., Hall, G. N., Rapley, J. et al. 2005, Plasma Physics and Controlled Fusion, 47, 465-B479
\bibitem[Lery \& Frank (2000)]{Lery00} Lery, T. \& Frank, A. 2000, \apj ,533, 897-910
\bibitem[Lim \& Raga (1998)]{Lim98} Lim, A. J. \& Raga, A. C. 1998, \mnras , 298, 871-876
\bibitem[Masciadri \& Raga (2001)]{Masciadri01} Masciadri, E. \& Raga, A. C. 2001, \aj , 121, 408-412
\bibitem[Michaut et al. (2004)]{Michaut04} Michaut, C., Stehlé, C., Leygnac, S., Lanz, T. \& Boireau, L. 2004, European Physical Journal D, 28, 381-392
\bibitem[Mitchell et al. (1996)]{Mitchell96} Mitchell, I. H., Bailey, J. M., Chittenden, J. P., Worley, J. F., Dangor, A. E., Haines, M. G. \& Choi, P. 1996, Review of Scientific Instruments, 67, 1533-1541
\bibitem[O'Dea \& Owen (1986)]{ODea86} O'Dea, C. P. \& Owen, F. N. 1986, \apj , 301, 841-859
\bibitem[Pavlov et al. (2003)]{Pavlov03} Pavlov, G. G., Teter, M. A., Kargaltsev, O. \& Sanwal, D. 2003, \apj , 591, 1157-1171
\bibitem[Raga et al. (2002)]{Raga02} Raga, A. C., de Gouveia Dal Pino, E. M., Noriega-Crespo, A., Mininni, P. D. \& Velázquez, P. F. 2002, \aap , 392, 267-276
\bibitem[Raga et al. (2007)]{Raga07} Raga, A. C., de Colle, F., Kajdic, P., Esquivel, A. \& Canto, J. 2007, \aap , 465, 879-885
\bibitem[Reipurth \& Bally (2001)]{Reipurth01} Reipurth, B. \& Bally, J. 2001, \araa , 39, 403-455
\bibitem[Ryutov et al. (1999)]{Ryutov99} Ryutov, D., Drake, R. P., Kane, J., Liang, E., Remington, B. A. \& Wood-Vasey, W. M. 1999, \apj , 518, 821-832
\bibitem[Stehle \& Chieze (2002)]{Stehle02} Stehle, C. \& Chieze, J.-P. 2002, in Semaine de l'Astrophysique Francaise, Conference Series, ed. F. Combes \& D. Barre, EdP-Sciences (Editions de Physique) 493
\end{thebibliography}
\end{document}